\documentclass{extarticle}
\usepackage[T1]{fontenc}
\usepackage[latin9]{inputenc}
\usepackage{amsmath}
\usepackage{amssymb}
\usepackage{cancel}
\usepackage{graphicx}
\usepackage{esint}

\makeatletter

\providecommand{\tabularnewline}{\\}

\makeatother

\begin{document}
\title{\textbf{\large{}Isovector Meson Masses from QCD Sum Rules}}
\author{\textbf{\normalsize{}Nasrallah F. Nasrallah$^{(a)}$, Karl Schilcher}\textbf{$^{(b,c)}$}\\
{\normalsize{}$^{(a)}$ Faculty of Science, Lebanese University, Tripoli
1300, Lebanon }\\
{\normalsize{}$^{(b)}$ Institut für Physik, Johannes Gutenberg-Universität}\\
{\normalsize{}Staudingerweg 7, D-55099 Mainz, Germany}\\
{\normalsize{}$^{(c)}$Centre for Theoretical Physics and Astrophysics}\\
{\normalsize{}University of Cape Town, Rondebosch 7700, South Africa}\\
\thanks{Supported in part by NRF (South Africa) and Alexander von Humbold
Foundation (Germany)}}
\maketitle
\begin{abstract}
\noindent We present a calculation of the masses of the isovector
mesons ( vector, scalar and pseudoscalar including the established
recurrences) using a new method of finite energy QCD sum rules. The
method is based on the idea of choosing a \ suitable integration
kernel which minimizes the occurring integral over the cut in the
complex energy (squared) plane. We obtain remarkably stable results
in a wide range $R$, where $R$ is the radius of the integration
contour. The sum rule predictions agree with the experimental values
within the expected accuracy showing that QCD describes single resonances.

\end{abstract}
\newpage{}

\section{QCD sum rules and meson masses}

The mass of the $\rho$-meson was first calculated from QCD and the
operator product expansion (OPE) in the pioneering paper of Shifman,
Vainshtein and Zakharov \cite{SVZ}. The calculation is based on QCD
sum-rules of the Borel (or Laplace) type. Although the results were
at the time rather spectacular, it was soon recognized \cite{Ioffe}
that, apart from the arbitrariness of the integration kernel, they
suffer from instabilities related to the specific choice of the Borel
variable and the assumption of a constant continuum of the isovector
spectral function. There are two arbitrary parameters, the Borel parameter,
called $M_{0}^{2}$, and the onset of the continuum. Note, in the
original paper the authors based their analysis on a small QCD coupling,
corresponding to a scale $\Lambda_{\text{QCD}}\approx100$ MeV compared
to the modern value $\Lambda_{\text{QCD}}\approx350$ MeV).

A more stringent approach is based on finite energy sum rules \cite{Kataev}.
We developed this idea further and used it in numerous applications
\cite{Dom}. Here we will use our approach to calculate all relevant
hadron masses, starting with the $\rho$-meson \cite{PDG} 
\[
m_{\rho}=0.775.25\pm.0.00026\text{ GeV\ }m_{\rho}^{2}=0.6010\text{ GeV}^{2}
\]

Consider the isovector current
\begin{equation}
j_{\mu}=\frac{1}{2}(\bar{u}\gamma_{\mu}u-\bar{d}\gamma_{\mu}d)\label{21}
\end{equation}
with the quantum numbers of the $\rho$. The relevant spectral function
is given by the absorptive part of the correlator
\begin{eqnarray}
\Pi_{\mu\nu}(q) & = & i\int d^{4}xe^{iqx}\left\langle 0\left\vert T\,j_{\mu}(x)j_{\nu}(0)\right\vert 0\right\rangle \label{22}\\
 & = & (q_{\mu}q_{\nu}-g_{\mu\nu}q^{2})\Pi(q^{2})\label{23}
\end{eqnarray}
Phenomenologically the spectral functions built up by the $\rho$
and the higher resonances $\rho^{\prime}(1450)$, $\rho^{\prime\prime}(1700)$,....
Neglecting its width, the experimental spectral function of the $\rho$
is given by 
\begin{equation}
\rho^{\text{exp}}(t)=\frac{m_{\rho}^{2}}{g_{\rho}^{2}}\delta(t-m_{\rho}^{2})=\frac{1}{\pi}\text{Im}\Pi^{\text{exp}}(t)\label{24}
\end{equation}
where $g_{\rho}$ is defined by
\[
\left\langle 0\left\vert j_{\mu}(0)\right\vert \rho(p,s)\right\rangle =\frac{m_{\rho}^{2}}{g_{\rho}}\varepsilon_{\mu}
\]
with $g_{\rho}=4.97\pm0.07$ as determined from the leptonic decay
of the rho-meson. The spectral function of Eq.(\ref{24}) corresponds
to an amplitude
\begin{equation}
\Pi^{\text{exp}}(t)=-\frac{m_{\rho}^{2}}{g_{\rho}^{2}}\frac{1}{(t-m_{\rho}^{2})}\text{.}\label{25}
\end{equation}
To lowest non-trivial order, the corresponding\ QCD expression is
\begin{equation}
\Pi^{\text{QCD}}(t)=-\frac{1}{8\pi^{2}}(1+a_{s})L+\frac{\left\langle m_{u}\bar{u}u+m_{d}\bar{d}d\right\rangle }{2t^{2}}+\frac{\left\langle a_{s}GG\right\rangle }{24t^{2}}+\frac{112\pi}{81t^{3}}(\sqrt{\alpha_{s}}\left\langle \bar{q}q\right\rangle )^{2}+...\label{26a}
\end{equation}
where $a_{s}=\alpha_{s}/\pi$ is the strong coupling at the scale
$\mu$ and$\ L\equiv\ln\frac{-t}{\mu^{2}}$. We take $a_{s}=0.1$
for $\mu$ of order $3\text{ GeV}^{2}$ to $4\text{ GeV}^{2}$ as
measured in $\tau$-decay. The variation of $a_{s}$ in this region
is of higher order.

To next order in QCD and the $\overline{MS}$ scheme, the perturbative
part of the vector correlator is given by \cite{Chetyrkin}
\begin{equation}
8\pi^{2}\Pi^{\text{QCD}}=-\left[1+L+aL+a^{2}(F_{3}L+\frac{\beta_{1}}{4}L^{2})\right]+...\label{26b}
\end{equation}
where
\[
\beta_{1}=-\frac{1}{2}(11-\frac{2}{3}n_{f}),\ \ \ \ F_{3}=1.9857-0.1153n_{f}
\]

The basis of FESR is Cauchy's theorem applied to the contour of Fig.
1

\begin{figure}[h]
\begin{centering}
\includegraphics[width=0.7\textwidth]{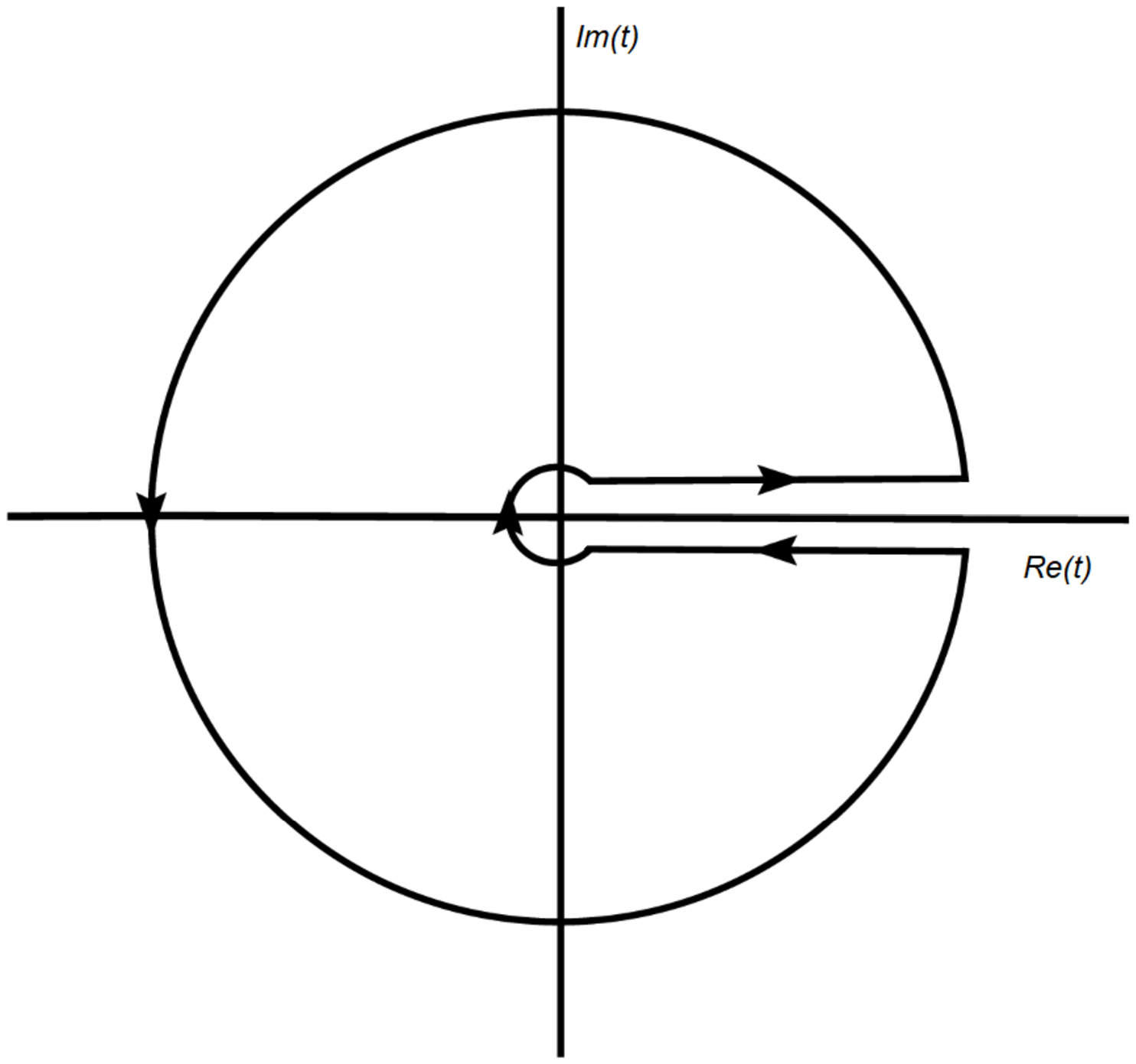} 
\par\end{centering}
\centering{}Fig. 1: Integration contour of FESR
\end{figure}

which implies
\begin{equation}
-\frac{m_{\rho}^{2}}{g_{\rho}^{2}}P(m_{\rho}^{2})=\frac{1}{\pi}\int\limits _{\text{cut}}^{R}dt\,P(t).Im\Pi^{\text{exp}}(t)+\frac{1}{2\pi i}\oint\limits _{\left\vert t\right\vert =R}dt\,P(t)\Pi^{\text{QCD}}(t)\label{27}
\end{equation}
where $P(t)$ is an entire function, e. g. a polynomial. Over the
circle of large radius $R$ the correlator $\Pi(t)$ has been replaced
by its QCD expression. The principal unknown in Eq.(\ref{27}) is
the integral over the cut, i. e. over the higher vector-isovector
resonances with mass $m_{\rho_{i}}^{2}\leq R$. To minimize this integral
(before neglecting it), a judicious choice of the weight-function
$P(t).$ has to be made. With the classic choice \cite{SVZ} $P(t)=\exp(-t/M_{0}^{2})$
the Borel variable $M_{0}^{2}$ cannot be chosen too large because
it would minimize the contribution of the\ $\rho$-meson. Also $M_{0}^{2}$
cannot be too small because the unknown condensates in Eq.(\ref{26a})
would explode. It was hoped in \cite{SVZ} that a region of stability
at an intermediate $M_{0}^{2}$ can be found. This can be shown to
be not the case \cite{Ioffe}. In our FESR approach we take $P(t)$
to be a polynomial
\begin{equation}
P(t).=\sum_{n=0}^{n_{\text{max}}}c_{n}t^{n}\text{ .}\label{28}
\end{equation}
It is clear that the order $n_{\text{max}}$ cannot be chosen arbitrarily
high because of the contribution of unknown condensates. We choose
a polynomial $P_{1}(t)$\ which vanishes at the mass $m_{1}=(1465\pm25)$
MeV \cite{PDG} ($m_{1}^{2}=2.15\text{ GeV}^{2}$) of the first resonance
and at the integration radius $R$. Explicitly, we take

\begin{eqnarray}
P_{1}(t,R). & = & (1-\frac{t}{m_{1}^{2}})(1-\frac{t}{R})\label{1.5}\\
 & = & 1-a_{1}t-a_{2}t^{2}\text{ \ where \ }a_{1}(R)=\frac{1}{m_{1}^{2}}+\frac{1}{R}\text{, \ }a_{2}(R)=-\frac{1}{m_{1}^{2}R}\label{1.5a}
\end{eqnarray}

When necessary we take $R\gtrsim3$ GeV$^{2}$ ($\sim m_{\tau}^{2}$)
as we know from the $\tau$-decay analysis \cite{Pich} that global
duality is valid there. Later we will need $P_{1}(m_{\rho}^{2}=0.6\text{ GeV}^{2}$,
$R=3\text{ GeV}^{-2}$$)=0.576$ and\ $a_{1}(R=3\text{ GeV}^{2}$)
$=0.799\,$\ GeV$^{-2}$ and $a_{2}(R=3\text{ GeV}^{2}$$)=$ $-0.155$
GeV$^{-4}$.

If we neglect the contribution of the physical continuum for $0\leq t\leq R$
and use Cauchy's theorem we arrive at the sum-rule 
\begin{eqnarray}
\frac{m_{\rho}^{2}}{g_{\rho}^{2}}P_{1}(m_{\rho}^{2},R) & = & \frac{1}{2\pi i}\oint\limits _{\left\vert t\right\vert =R}dt\,P_{1}(t,R)\Pi^{\text{QCD}}(t)\nonumber \\
 & = & \frac{1}{8\pi^{2}}(1+a_{s})\int\limits _{0}^{R}dt\,P_{1}(t,R).-a_{1}\frac{\left\langle m_{u}\bar{u}u+m_{d}\bar{d}d\right\rangle }{2}-a_{1}\frac{\left\langle a_{s}GG\right\rangle }{24}+....\label{1.6}
\end{eqnarray}
As a check of our method we plot the integral appearing in Eq.(\ref{1.6}).
As can be seen from Fig. 2.

\begin{equation}
I_{0}(R)=\int_{0}^{R}(1-\frac{t}{m_{1}^{2}})(1-\frac{t}{R})dt=\frac{R}{2}(1-\frac{1}{3}\frac{R}{m_{1}^{2}})\text{ }\label{34c}
\end{equation}
\begin{figure}[h]
\begin{centering}
\includegraphics[width=0.7\textwidth]{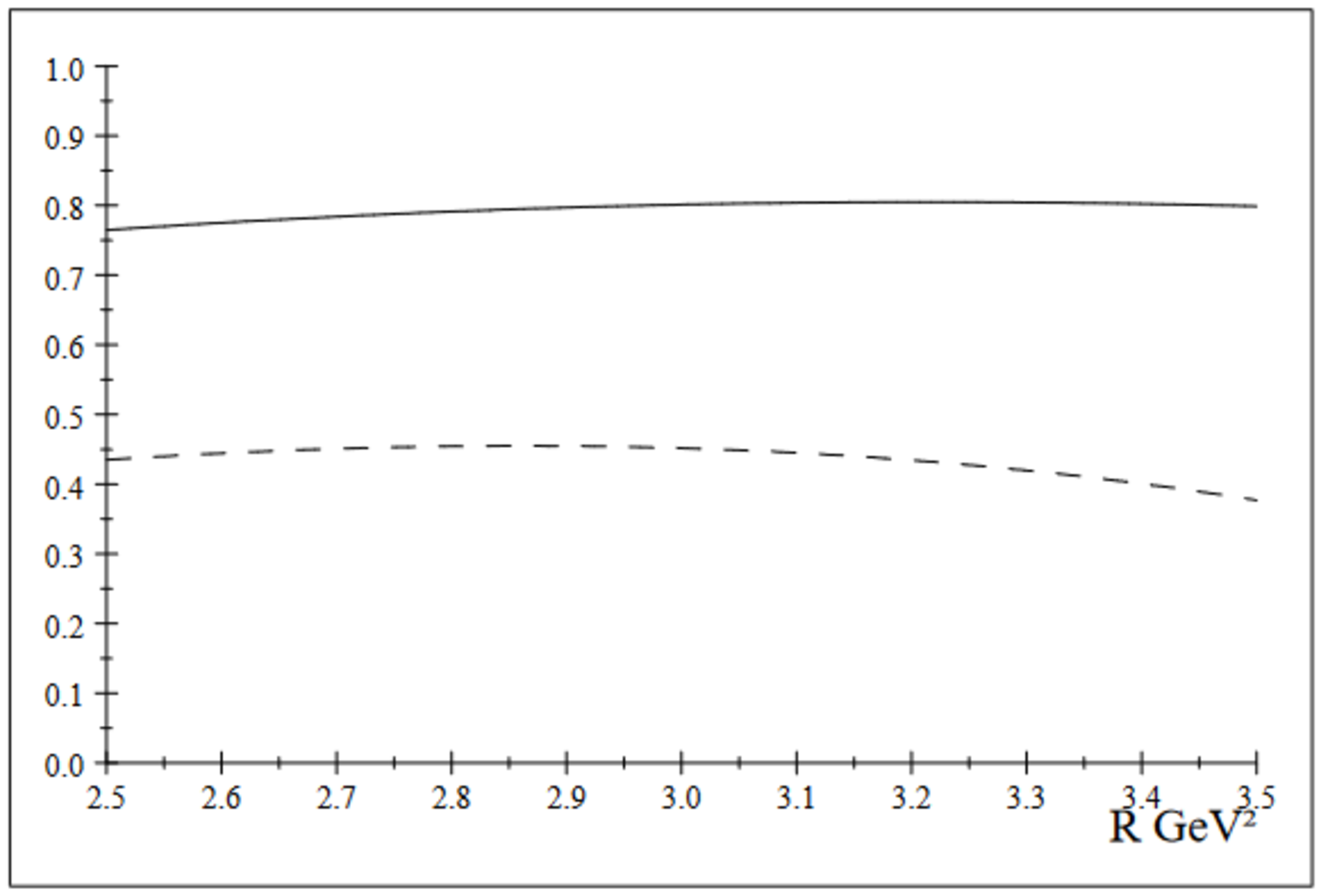} 
\par\end{centering}
\centering{}Fig. 2: The integrals $I_{0}(R)$ (solid line) and $I_{1}(R)$
(dashed line)
\end{figure}

If we look for stability of the result for $m_{\rho}$ we have to
restrict $2.5\text{ GeV}^{2}$$\leq R\leq3.5\text{ GeV}^{2}$, a typical
result being $I_{0}(3GeV^{2}$$)=0.801\text{ GeV}^{2}$. On the one
hand this is just the region where we expect perturbative QCD to be
valid, on the other hand we do not expect higher resonances to contribute
significantly to the sum rule in this region. Our results do not depend
on the precise choice of $R$ as an be seen from Fig. 2 as long as
$2.5\text{ GeV}^{2}$ $\leq R\leq3.5\text{ GeV}^{2}$. We will use
Eq.(\ref{1.6}) in this region in the chiral limit

\begin{eqnarray}
\frac{m_{\rho}^{2}}{g_{\rho}^{2}}P_{1}(m_{\rho}^{2},R) & = & \frac{1}{8\pi^{2}}(1+a_{s})\int\limits _{0}^{R}dt\,P_{1}(t,R).-\frac{a_{1}}{24}\left\langle a_{s}GG\right\rangle \label{1.6cl}\\
\frac{m_{\rho}^{2}}{g_{\rho}^{2}}0.583 & = & \frac{1.1}{8\pi^{2}}0.801\text{ GeV}^{2}-\frac{0.799\text{ GeV}^{-2}}{24}\left\langle a_{s}GG\right\rangle \label{1.7b}\\
 &  & \text{for }R\sim3\text{ GeV}^{2}\nonumber 
\end{eqnarray}
Experimentally, $g_{\rho}$ can be determined from its leptonic decay
width

\[
\Gamma(\rho\rightarrow e_{+}e_{-})=\frac{1}{3}\alpha^{2}m_{\rho}\frac{4\pi}{g_{\rho}^{2}}=7.04\pm0.06\text{ KeV}
\]

\begin{equation}
\frac{g_{\rho}^{2}}{4\pi}=1.96\pm0.02\label{1.6b}
\end{equation}
Neglecting the gluon condensate\ and setting $m_{\rho}^{2}=0.6\text{ GeV}^{2}$,
Eq. (\ref{1.7b}) gives
\[
\frac{g_{\rho}^{2}}{4\pi}=\frac{0.6\text{ GeV}^{2}}{4\pi}\left(\frac{1}{0.583}\frac{1.1\times0.801\text{ GeV}^{2}}{8\pi^{2}}\right)^{-1}=2.494
\]
This result is consistent with the experimental one Eq.(\ref{1.6b})
considering that the former involves the difference of two large numbers.
The error cannot be estimated reliably as it arises mainly from the
Ansatz of a narrow $\rho$-resonance for the spectral function.

The error due to $\Pi_{\text{pert}}^{\text{QCD}}$ is small. To order
$a^{2}$ ($=(\alpha_{s}/\pi)^{2}$) the correlator is given in Eq.(\ref{26b}).
The relevant integrals can be found in \cite{DS}.The error due to
neglected higher order perturbative terms turns out to be of order
2\% to 4\%, significantly smaller than the further errors to be discussed
below.

\section{The Isovector Vector Mesons}

We propose a sum rule method that is optimally suited to calculate
all resonance masses from QCD. We define a polynomial $P_{i}(t)$
\[
P_{i}(t)=\left(1-\frac{t}{m_{i}^{2}}\right)\left(1-\frac{t}{R}\right)
\]
\ which vanishes at the mass $m_{i}$ and at the integration radius
$R$. For example $m_{1}=(1465\pm25)$ MeV \cite{PDG} ($m_{1}^{2}=2.15\text{ GeV}^{2}$
) is the first resonance recurrence. In order to get the mass of the
$\rho$-meson we take the first moment integral 
\begin{eqnarray}
\frac{m_{\rho}^{4}}{g_{\rho}^{2}}P_{1}(m_{\rho}^{2},R) & = & \frac{1}{2\pi i}\oint\limits _{\left\vert t\right\vert =R}dt\,t\,P_{1}(t,R).\Pi^{\text{QCD}}(t)\nonumber \\
 & = & \frac{1}{8\pi^{2}}(1+a_{s})\int\limits _{0}^{R}dt\,t\,P_{1}(t,R).-\frac{\left\langle m_{u}\bar{u}u+m_{d}\bar{d}d\right\rangle }{2}-\frac{\left\langle a_{s}GG\right\rangle }{24}+....\label{1.7}
\end{eqnarray}

Consider the integral
\begin{equation}
I_{1}(R,m_{1}^{2})=\int_{0}^{R}(1-\frac{t}{m_{1}^{2}})(1-\frac{t}{R})\,t\,dt=\frac{R^{2}}{12}(2-\frac{R}{m_{1}^{2}})\label{1.7c}
\end{equation}
For $R=3.0\text{ GeV}^{2}$ and $m_{1}=1.465$ GeV the result is $I_{1}(3\text{ GeV}^{2}$$)=0.451\text{ GeV}^{4}$.
The result is still stable for $2.5\text{ GeV}^{2}$ $\leq R\leq3.5\text{ GeV}^{2}$,
see Fig.2. With the standard values of the condensates 
\[
\left\langle a_{s}GG\right\rangle =0.013\ \text{GeV}^{4}\ \ ,\ \ \left\langle m_{u}\bar{u}u+m_{d}\bar{d}d\right\rangle =-1.67\times10^{-4}\text{ GeV}^{4}
\]
and choosing $R=3.0\text{ GeV}^{2}$ Eqs.(\ref{1.6}) and (\ref{1.7})
give
\[
m_{\rho}=0.73\text{ GeV}
\]

Our choice (Eq.\ref{1.5}) for $P_{1}(t)$ provides a good damping
for the contribution of the continuum in the interval $2.5\text{ GeV}^{2}$$\leq t\leq3.5\text{ GeV}^{2}$:
The contribution of the resonances $\rho(1450)$ and $\rho(1700)$
(almost) vanishes and that of the $\rho(1580)$ is shrunk by a factor
of $P_{1}(1.58^{2}\text{ GeV}^{2})$ $/P_{1}(0.6\text{ GeV}^{2}$$)=-5.7\times10^{-2}$.
This renders the contribution of the continuum negligible.

It is nevertheless worthwhile to assess the influence\ of the variation
of $P(t)$ in the result for $m_{\rho}$ in order to estimate the
error inherent in the method.

One choice would be
\[
P(t)=\left(1-\frac{t}{2.1\text{ GeV}^{2}}\right)\left(1-\frac{t}{2.5\text{ GeV}^{2}}\right)\left(1-\frac{t}{2.89\text{ GeV}^{2}}\right)
\]
yielding $m_{\rho}=0.71$ GeV. Or
\[
P(t)=\left(1-\frac{t}{2.5\text{ GeV}^{2}}\right)^{2}
\]
which gives $m_{\rho}=0.79$ GeV. We have tried several other polynomials,
so we give finally
\begin{equation}
m_{\rho}=(0.74\pm0.04)\text{ GeV}\label{1.8a}
\end{equation}

Taking an additional moment yields the mass $m_{1}$ of the $\rho_{1}(1450)$.
In addition to Eqs.(\ref{1.6}) and (\ref{1.7}) we have, neglecting
the higher condensates
\begin{equation}
\frac{m_{\rho}^{6}}{g_{\rho}^{2}}P_{1}(m_{\rho}^{2},R)=\frac{(1+a_{s})}{8\pi^{2}}\int_{0}^{R}dt\,t^{2}P_{1}(t,R)\label{1.13}
\end{equation}

The mass $m_{1}$ can be determined by the two ratios
\begin{eqnarray*}
m_{\rho}^{2} & = & \frac{\text{rhs of Eq.}(\ref{1.13})\text{ }}{\text{rhs of Eq.}(\ref{1.7})}=\frac{\int_{0}^{R=3GeV^{2}}dt\,t^{2}P_{1}(t,R)}{\int_{0}^{R=3GeV^{2}}dt\,tP_{1}(t,R)}\\
 & = & \frac{\text{rhs of Eq.}(\ref{1.7})\text{ }}{\text{rhs of Eq.}(\ref{1.6})}=\frac{\int_{0}^{R=3GeV^{2}}dt\,tP_{1}(t,R)}{\int_{0}^{R=3GeV^{2}}dt\,P_{1}(t,R)}
\end{eqnarray*}
all at $R=3.0\text{ GeV}^{2}$. The mass $m_{1}$ is determined by
the equating the two ratios using Eqs. (\ref{34c}), (\ref{1.7c})
and

\begin{equation}
I_{2}(R,m_{1}^{2})=\int_{0}^{R}(1-\frac{t}{m_{1}^{2}})(1-\frac{t}{R})t^{2}dt=\frac{R^{3}}{60}(5-3\frac{R}{m_{1}^{2}})\label{1.7d}
\end{equation}
From
\[
\frac{I_{1}}{I_{0}}=\frac{I_{2}}{I_{1}}
\]
we obtain 
\[
m_{1}=m_{\rho}(1450)=1.42\pm0.10\text{ GeV}
\]
The error is again estimated by varying $R$ by $10\%$.

\begin{figure}[h]
\begin{centering}
\includegraphics[width=0.7\textwidth]{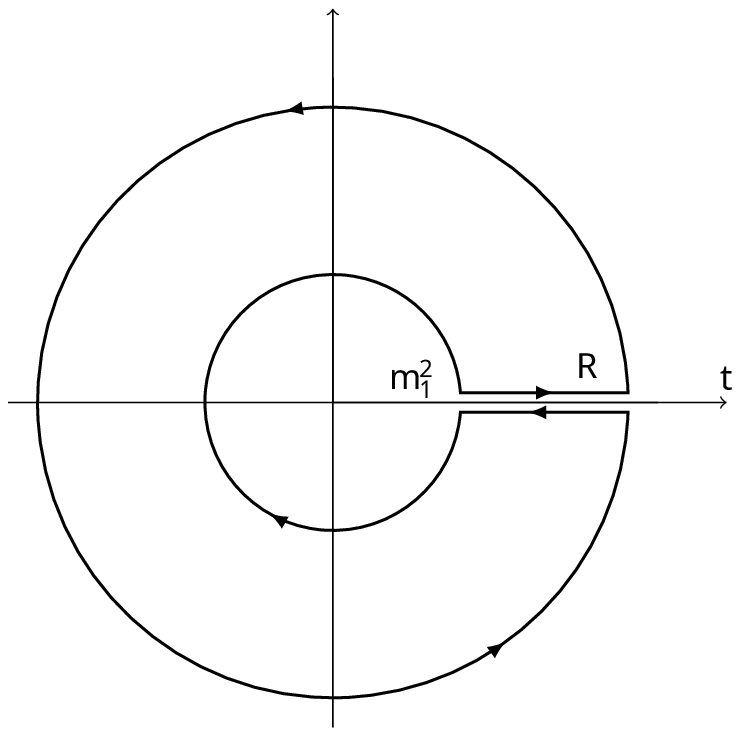} 
\par\end{centering}
\centering{}Figure 3: Alternative integration contour
\end{figure}

We can proceed further and consider the integral over the contour
indicated in Fig. 3 with the kernel 
\begin{equation}
P_{3}(t,R)=\left(1-\frac{t}{m_{3}^{2}}\right)\left(1-\frac{t}{R}\right)\label{1.15}
\end{equation}

In a calculation of $m_{i}$ we choose $m_{i-1}^{2}$ as the lower
limit of integration because the contour starts there. In addition
we have a check of the consitency of the full set of isovector meson
mass determination. $R$ (and $a_{s})$ is again chosen in the stability
region ($R\sim3.5\text{ GeV}^{2}$).

We have to assume here that global duality is for $t\geq m_{1}^{2}=2.15\text{ GeV}^{2}$.

Neglecting all condensates the mass of the $\rho_{2}=\rho(1580)$
is obtained from the sum rules
\begin{eqnarray}
\frac{m_{2}^{2}}{g_{\rho_{2}}^{2}}P_{3}(m_{2}^{2},R & = & 3.5)=\frac{(1+a_{s})}{8\pi^{2}}\int_{m_{1}^{2}}^{3.5\text{ GeV}^{2}}dt\,P_{3}(t,R)\label{1.16}\\
\frac{m_{2}^{4}}{g_{\rho_{2}}^{2}}P_{3}(m_{2}^{2},R & = & 3.5)=\frac{(1+a_{s})}{8\pi^{2}}\int_{m_{1}^{2}}^{3.5\text{ GeV}^{2}}dt\,tP_{3}(t,R)\label{1.17}
\end{eqnarray}
taking the mass $m_{3}=m_{\rho}(1700)$ (entering via $P_{3}$) as
given. The ratio of the above integrals gives
\begin{equation}
m_{2}^{2}=\frac{\int_{m_{1}^{2}}^{3.5\text{ GeV}^{2}}dt\,tP_{3}(t,R)}{\int_{m_{1}^{2}}^{3.5\text{ GeV}^{2}}dt\,P_{3}(t,R)}\label{1.18}
\end{equation}
in the stability region ($R\sim3.5\text{ GeV}^{2}$)
\begin{equation}
m_{2}=m_{\rho}(1580)=(1.50\pm0.05)\text{ GeV}\label{1.18c}
\end{equation}
The error is obtained by varying $R$ by $\pm10\%$.

Similarly we can obtain $m_{3}=m_{\rho}(1700)$ from the kernel
\[
P_{4}(t,R)=\left(1-\frac{t}{m_{4}^{2}}\right)\left(1-\frac{t}{R}\right)
\]
and integrating over the contour of Fig. 3. Assuming QCD duality is
valid from $m_{2}^{2}=1.465^{2}=2.146\text{ GeV}^{2}$the corresponding
sum rule reads
\begin{equation}
m_{3}^{2}=\frac{\int_{m_{2}^{2}}^{R}dt\,tP_{4}(t,R)}{\int_{m_{2}^{2}}^{R}dt\,P_{4}(t,R)}\label{1.19a}
\end{equation}

For $m_{4}=2.150$ GeV the stability region being $R\sim4.5\text{ GeV}^{2}$.
The result is
\[
m_{3}=1.71\text{ GeV}
\]
The mass $m_{3}$ can also be obtained from $P_{3}(t,R=3.5\text{ GeV}^{2}$$)=\left(1-\frac{t}{m_{3}^{2}}\right)\left(1-\frac{t}{R}\right)$
from 
\[
m_{2}^{2}=\frac{\int_{m_{1}^{2}}^{R=3.5\text{ GeV}^{2}}dt\,t^{2}P_{3}(t,R)}{\int_{m_{1}^{2}}^{R=3.5\text{ GeV}^{2}}dt\,tP_{3}(t,R)}=\frac{\int_{m_{1}^{2}}^{R=3.5\text{ GeV}^{2}}dt\,tP_{3}(t,R)}{\int_{m_{1}^{2}}^{R=3.5\text{ GeV}^{2}}dt\,P_{3}(t,R)}
\]
The equality of the ratios gives
\begin{equation}
m_{3}=1.74\text{ GeV}\label{1.22}
\end{equation}

Combining the two results we obtain
\[
m_{3}=m_{\rho}(1700)=1.73\pm0.05\text{ GeV}
\]
We finally proceed to calculate $m_{4}=m_{\rho}(2150)$ using the
kernel
\[
P_{4}(t,R)=\left(1-\frac{t}{m_{4}^{2}}\right)\left(1-\frac{t}{R}\right)
\]
in the stability region $R\sim6.0\text{ GeV}^{2}$. The mass $m_{4}$
follows from the sum rule
\begin{equation}
m_{3}^{2}=\frac{\int_{m_{2}^{2}}^{6.0\text{ GeV}^{2}}dt\,tP_{4}(t,R)}{\int_{m_{2}^{2}}^{6.0\text{ GeV}^{2}}dt\,P_{4}(t,R)}=\frac{\int_{m_{2}^{2}}^{6.0\text{ GeV}^{2}}dt\,t^{2}P_{4}(t,R)}{\int_{m_{2}^{2}}^{6.0\text{ GeV}^{2}}dt\,tP_{4}(t,R)}\text{ \ all at }R=6.0\text{ GeV}^{2}
\end{equation}

Equality of the ratios give
\[
m_{4}=m_{\rho}(2150)=2.18\pm0.09\text{ GeV}
\]

The error is obtained by varying $R$ by 10\%.

We collect our results together with the experimental numbers from
\cite{PDG} in a table. 

\textbf{Table\ of Results}:
\begin{center}
\begin{tabular}{ccc}
Resonance  & Result for the mass in GeV  & Experimental value in GeV \tabularnewline
$\rho(770)$  & $0.74\pm0.04$  & $0.77511\pm0.00034$ \tabularnewline
$\rho(1450)$  & $1.42\pm0.10$  & $1.465\pm0.025$ \tabularnewline
$\rho(1570)$  & $1.50\pm0.05$  & $1.570\pm0.036\pm0.062$ \tabularnewline
$\rho(1700)$  & $1.73\pm0.05$  & $1.720\pm0.020$ \tabularnewline
$\rho(2150)$  & $2.18\pm0.09$  & $2.155\pm0.021$\tabularnewline
\end{tabular}
\par\end{center}

We conclude that the QCD sum rules can predict the masses of all established
higher\ $\rho$ recurrences. QCD describes in this case a single
resonance.

\section{The Isovector Pseudoscalars}

The following isovector pseudoscalars have been observed $\pi$ with
$m_{\pi}\approx0$ GeV, $\pi_{1}(1300)$ with $m_{1}=1.3$ $\pm0.1$
GeV, $\pi_{2}(1810)$ with $m_{2}=$ $1.81\pm0.01$ GeV, $\pi_{3}(2370)$
with $m_{3}=2.360\pm25$ GeV \cite{PDG}.

We start with the correlator
\begin{equation}
\Pi(q)=i\int d^{4}xe^{iqx}\left\langle 0\left\vert T\,j_{5}(x)j_{5}(0)\right\vert 0\right\rangle \label{2.1}
\end{equation}
of the pseudoscalar current 
\begin{equation}
j_{5}=i\overline{q}\gamma_{5}q\qquad j_{5}=\frac{1}{2m_{q}}\partial^{\mu}A_{\mu}\label{2.1a}
\end{equation}
where $m_{q}$ is the quark mass, $q=u$ or $d$. The QCD expression
for the correlator is
\begin{equation}
\Pi^{\text{QCD}}(t)=-\frac{3}{8\pi^{2}}(1+\frac{11}{3}a_{s})t\ln(-t)+\frac{\left\langle m_{q}^{2}\overline{q}q\right\rangle }{t}+\frac{\left\langle a_{s}GG\right\rangle }{8t}+...\label{2.2}
\end{equation}
The scale $\mu^{2}$ only enters at order $\alpha_{s}^{2}$. The method
used for the vector mesons is repeated here. We use the kernel $P_{2}(t,R).=(1-\frac{t}{m_{2}^{2}})(1-\frac{t}{R})$
which vanishes at $t=m_{2}^{2}$ and $t=R$ to get the sum rule

\begin{eqnarray*}
m_{1}^{2} & = & \frac{1}{8\pi^{2}}(1+\frac{11}{3}a_{s})\int\limits _{0}^{R}dt\,t^{2}\,P_{2}(t,R)\\
 & = & \frac{\int_{0}^{\text{R }}dt\,t^{2}P_{2}(t,R)}{\int_{0}^{\text{R }}dt\,tP_{2}(t,R)-\delta}
\end{eqnarray*}
where

\[
\delta=\frac{\pi^{2}\left\langle a_{s}GG\right\rangle }{(1+\frac{11}{3}a_{s})}=0.094\text{ GeV}^{4}
\]
With the stability region $R\approx4.0\text{ GeV}^{2}$ this gives
\begin{equation}
m_{1}=m_{\pi}(1300)=1.22~\text{GeV .}\label{2.6}
\end{equation}
Taking an additional moment with the kernel
\begin{equation}
P_{2}(t,m_{2})=\left(1-\frac{t}{m_{2}^{2}}\right)\left(1-\frac{t}{R}\right)\text{ \ at }R=4.2\text{ GeV}^{2},\label{2.6a}
\end{equation}
we obtain the consistency condition
\begin{equation}
\frac{\int_{0}^{\text{R }}dt\,t^{3}P_{2}(t,R)}{\int_{0}^{\text{R }}dt\,t^{2}P_{2}(t,R)}=\frac{\int_{0}^{\text{R }}dt\,t^{2}P_{2}(t,R)}{\int_{0}^{\text{R }}dt\,t\,P_{2}(t,R)-\delta}\text{, \ all at }R=4.2\text{ GeV}^{2}\label{2.7}
\end{equation}
This yields
\[
m_{2}=m_{\pi}(1810)=1.77\text{ GeV}
\]

Alternatively one can use the kernel
\begin{equation}
P_{3}(t,R)=\left(1-\frac{t}{m_{3}^{2}}\right)\left(1-\frac{t}{R}\right)\label{2.10}
\end{equation}
in the sum rule
\begin{equation}
m_{2}^{2}=\frac{\int_{1.69}^{\text{R }}dt\,t^{2}P_{3}(t,R)}{\int_{1.69}^{\text{R }}dt\,t\,P_{3}(t,R)-\delta}\text{ .}\label{2.11}
\end{equation}
assuming optimistically that QCD duality is valid from $R=1.69\text{ GeV}^{2}$.
At stability ($R\approx7\text{ GeV}^{2}$) this gives
\begin{equation}
m_{2}=m_{\pi}(1810)=1.74\text{ GeV}\label{2.12}
\end{equation}
One can calculate the mass $m_{3}$ making use of the kernel $P_{3}(t,m_{2})$
at $R=4.4\text{ GeV}^{2}$,and imposing the condition
\begin{equation}
\frac{\int_{m_{2}^{2}}^{7.1\text{ GeV}^{2}}dt\,t^{3}P_{3}(t,m_{3})}{\int_{m_{2}^{2}}^{7.1\text{ GeV}^{2}}dt\,t^{2}P_{3}(t,m_{3})}=\frac{\int_{m_{2}^{2}}^{7.1\text{ GeV}^{2}}dt\,t^{2}P_{3}(t,m_{3})}{\int_{m_{2}^{2}}^{7.1\text{ GeV}^{2}}dt\,P_{3}(t,m_{3})-\delta}\label{2.14}
\end{equation}
yields
\[
m_{3}=m_{X}(2370)=2.66\text{ GeV}
\]
This is an argument for the isovector pseudoscalar nature of the X(2370).

\textbf{Table\ of Results}:

\[
\begin{tabular}{ccc}
 Resonance  &  Result for the mass in GeV  &  Experimental value in GeV \\
 \ensuremath{\pi_{1}(1300)}  &  \ensuremath{1.22}  &  \ensuremath{1.300} \\
 \ensuremath{\pi_{2}(1810)}  &  \ensuremath{1.77\pm0.04}  &  \ensuremath{1.810} \\
 \ensuremath{\pi_{3}=\pi_{3}(2370)}  &  \ensuremath{2.66}  &  \ensuremath{2.370} 
\end{tabular}
\]

The predictions, although qualitatively in agreement with the data,
are not as good as in the vector meson case mainly because QCD perturbation
theory is less convergent.

\section{The isovector scalar mesons}

The spectrum of the scalar mesons is $a_{0}(980)$, $a_{0}(1450)$,
$X(1835)$, $a_{0}(1950)$ where the status of the $X(1835)$ is uncertain.
We start with the correlator Eq.(\ref{2.1}) with the scalar current
\begin{equation}
j(x)=\bar{q}(x)q(x)\label{2.16}
\end{equation}
The QCD expression is the same as given in Eq.(\ref{2.2}) except
for the negligible $m_{q}^{2}\left\langle \bar{q}q\right\rangle $
term. The method gives
\begin{equation}
f_{1}^{2}P_{2}(m_{1}^{2},R)=\frac{1}{8\pi^{2}}(1+\frac{11}{3}a_{s})\int\limits _{0}^{R}dt\,t\,P_{2}(t,R)-\frac{\left\langle a_{s}GG\right\rangle }{8}+....\label{2.17}
\end{equation}
and
\begin{equation}
f_{1}^{2}m_{1}^{2}P_{2}(m_{1}^{2},R)=\frac{1}{8\pi^{2}}(1+\frac{11}{3}a_{s})\int\limits _{0}^{R}dt\,t^{2}\,P_{2}(t,R)-\frac{\left\langle a_{s}GG\right\rangle }{8}+....\label{2.18}
\end{equation}
with $P_{2}(m_{1}^{2},R)$ given by Eq. (\ref{2.6a}) with the input
$m_{2}^{2}=2.10\text{ GeV}^{2}$. At stability ($R\sim2.7\text{ GeV}^{2}$)
the ratio of the above equations gives
\begin{equation}
m_{1}=m_{a_{0}}(980)=0.93\text{ GeV}\label{2.19}
\end{equation}

We emphasize that we assume that the correlator is given by QCD perturbation
theory is for $|t|\,\geq R=2.7\text{ GeV}^{2}$as phenomenologically
supported by the analysis of $\tau$-decay \cite{Pich}.

Taking an additional moment with the kernel $P_{2}(t,R)=(1-\frac{t}{m_{2}^{2}})(1-\frac{t}{2.7GeV^{2}})$
and the condition
\begin{equation}
m_{1}^{2}=\frac{\int_{0}^{\text{R }}dt\,t^{3}P_{2}(t,R)}{\int_{0}^{\text{R }}dt\,t^{2}P_{2}(t,R)}=\frac{\int_{0}^{\text{R }}dt\,t^{2}P_{2}(t,R)}{\int_{0}^{\text{R }}dt\,t\,P_{2}(t,R)-\delta}\text{, \ all at }R=2.7\text{ GeV}^{2}\label{2.20a}
\end{equation}
Equating the two ratios yields
\[
m_{2}=m_{a_{0}}(1450)=1.51\text{ GeV}
\]

We next use the kernel $P_{4}(t,R)=(1-\frac{t}{m_{4}^{2}})(1-\frac{t}{R})$
\ with $m_{4}=m_{a_{0}}(1950)$ to determine $m_{3}$:
\begin{equation}
m_{3}^{2}=\frac{\int_{m_{2}^{2}}^{\text{R }}dt\,t^{2}P_{4}(t,R)}{\int_{m_{2}^{2}}^{\text{R }}dt\,t\,P_{4}(t,R)-\delta}\text{, \ all at }R=4.7\text{ GeV}^{2}\label{2.25}
\end{equation}
The result is $m_{3}=m_{X}(1835)=1.80$ GeV.

$m_{4}$ can be determined by
\begin{equation}
\frac{\int_{m_{2}^{2}}^{\text{R }}dt\,t^{3}P_{4}(t,R)}{\int_{m_{2}^{2}}^{\text{R }}dt\,t^{2}P_{4}(t,R)}=\frac{\int_{m_{2}^{2}}^{\text{R }}dt\,t^{2}P_{4}(t,R)}{\int_{m_{2}^{2}}^{\text{R }}dt\,t\,P_{4}(t,R)-\delta}\text{, \ all at }R=4.7\text{ GeV}^{2}
\end{equation}
with the result
\[
m_{4}=m_{a_{0}}(1950)=1.80\text{ GeV}
\]
\pagebreak{}

\textbf{Table of results}:
\begin{center}
\begin{tabular}{ccc}
Resonance & Result for the mass in GeV & Experimental value in GeV\tabularnewline
$\ensuremath{a_{0}(980)}$ & $0.93$ & $0.98$\tabularnewline
$a_{0}(1450)$ & $1.51$ & $1.45$\tabularnewline
$X(1835)$ & $1.80$ & $1.895$\tabularnewline
$a_{0}(1950)$ & $1.80$ & $1.930$\tabularnewline
\end{tabular}
\par\end{center}

\textbf{Conclusions}: We have calculated the masses of the isovector
(vector, pseudoscalar, scalar) mesons and their recurrences with a
new variant of QCD finite energy sum rules. The method works well
for all similar systems such as the nucleon resonances. The main source
of error is the zero width approximation for the resonances. We have
estimated this error by allowing the radius entering the sum rule
to vary by $\pm10\%$. Order $\alpha_{s}$ corrections are included,
order $\alpha_{s}^{2}$ are calculated and found to be negligible.
The sum rule \ predictions are compared with the experimental numbers
and agreement within the expected accuracy is found. It can \ be
concluded that QCD is applicable to single resonances and their recurrences.

\textbf{Acknowledgements:} This work was supported in part by the
Alexander von Humboldt Foundation (Germany), under the Research Group
Linkage Programme, and by the University of Cape Town (South Africa).

\end{document}